\documentstyle[11pt,aaspp]{article}

\begin{document}

\title{Why Disks Shine: the Transport of Angular Momentum in Hot, Thin Disks}
\author{E.T. Vishniac}
\affil{Department of Astronomy, University of Texas, Austin, TX 78712}

\begin{abstract}
I review recent work on the radial transport of angular momentum
in ionized, Keplerian accretion disks.  Proposed mechanisms
include hydrodynamic and MHD local instabilities and long range
effects mediated by wave transport.  The most promising models
incorporate the Velikhov-Chandrasekhar instability,
caused by an instability of the magnetic field embedded in
a differentially rotating disk.  This has the
important feature that the induced turbulent motions necessarily
transport angular momentum outward.  By contrast, convective modes
may transport angular momentum in either direction.  Combining the
magnetic field instability with an $\alpha-\Omega$ dynamo driven
by internal waves leads to a model in which the dimensionless
viscosity scales as $(H/r)^{4/3}$.  However, this model has a
phenomenology which is quite different from the $\alpha$ disk
model.  For example, an active disk implies some source of
excitation for the internal waves.  In binary systems with a mass
ratio of order unity the most likely exciting mechanism is a
parametric instability due to tidal forces.  This implies that in systems
where the accretion stream is intermittent, like MV Lyrae or TT
Ari, epochs when the mass flow is absent or very small will be
epochs in which the disk shrinks and becomes relatively inactive
and dark.  This model also implies that forced vertical mixing is
important, even in convectively stable disks.  I discuss various
observational tests of this model and the focus of current theoretical work.
\end{abstract}

\keywords{accretion, binaries}

\section{Introduction}

One of the outstanding problems in the theory of interacting binary
stars, in fact in modern astrophysical theory, is the question of
how angular momentum is transported in accretion disks.
This cannot be explained by microscopic viscosity, i.e.
atoms carrying momentum around.  It must be some sort of collective
effect.  The fundamental physical forces involved are all well understood.
Nevertheless, this is a difficult problem for at least three reasons.
First, the relevant physics is probably very complicated, defying
any straightforward analytic treatment. Second, the relevant
computer models are always grossly simplified. No one runs high
resolution three dimensional magnetohydrodynamics simulations
with realistic radiative transfer.  Third, the observational
constraints are hard to apply.  Partly because unambiguous
theoretical predictions are so hard to come by, and partly because
those aspects of an accretion disk that present themselves to
an observer are often only tangentially related to the essential
physics of these objects.
The traditional solution is to invoke a dimensional estimate, i.e.
\begin{equation}
\nu=\alpha c_s H
\end{equation}
as suggested by Shakura \& Sunyaev (1973).  Here $H$ is the disk thickness,
$c_s$ is the average sound speed, and $\alpha$ is (repeatedly!) adjusted
according to the demands of the observations.
Curiousity about the underlying physics, and the apparent plasticity of
$\alpha$ drive continued work in this area.

In the work presented here my collaborators and I have employed several
simplifying assumptions.  First, we assume we are considering geometrically
thin
disks, i.e.  $H\ll r$, where $r$ is the disk radius.  With this
assumption it becomes possible to use $H/r$ as an ordering parameter.
Second, we consider only disks with negligible self-gravity.  Both
the radial and vertical gravitational forces within the disk are
assumed to be dominated by the gravity of the central star,
which implies that $ \Omega(r)\propto r^{-3/2}$, where $\Omega$ is the
orbital frequency as a function of radius, $c_s\sim H\Omega$,
and vertical gravity is $-z\Omega^2$.  Third, we will only consider
disks composed of perfectly conducting gas.  This excludes large
regions within protostellar disks.  Finally, we will neglect all
external magnetic fields.  Each disk generates its
own internal field.  Elsewhere we have argued that this internal
field dominates the dynamics of the disk except under exceptional
circumstances.  Each of these conditions need not apply everywhere
in a disk.  The ideas presented here will apply to those parts of
those disks where these conditions are satisfied.

In what follows I will identify the critical processes which are
important for angular momentum transport in disks, and briefly
discuss why others are not.  I will then present the internal
wave driven dynamo model for angular momentum transport in accretion
disks and discuss some of the implications of this model.  Most
of the points discussed here are explained at length elsewhere
(Vishniac \& Diamond 1989, Vishniac, Jin, \& Diamond 1990,
Vishniac \& Diamond 1992, Vishniac \& Diamond 1993).

\section{Perturbation Modes in an Accretion Disk}

Our discussion will be based on a consideration of the local modes
that might reasonably be found in a magnetized accretion disk.
I will assume that the large scale magnetic field is largely
azimuthal, due to the strong shearing in the disk.
Two families of disk perturbations are closely related to
hydrodynamic modes, i.e. sound waves and internal waves
(or g-modes and p-modes to stellar structure theorists).
The former have the dispersion relation
\begin{equation}
\bar\omega^2\approx c_s^2k^2,
\end{equation}
where $\bar\omega=\omega+m\Omega$ is the frequency measured by an
observer moving in a circular orbit with frequency $\Omega(r)$.
The radial propagation of these waves is severely hampered by
refraction effects.  They tend to bend towards regions of low sound speed,
like the disk atmosphere, and steepen into weak shocks
which rapidly dissipate.  They may contribute to the heating of the
disk corona, but are unlikely to play a major role in angular momentum
transport.

Internal waves have the dispersion relation
\begin{equation}
k_r^2+\left({m\over r}\right)^2\approx
k_z^2\left({\Omega^2-\bar\omega^2\over\bar\omega^2-N^2}\right),
\label{eq:disp}
\end{equation}
where $N^2\equiv z\Omega^2{d\over dz}[\ln(P^{1/\gamma}/\rho)]$ is the
vertical buoyancy frequency.  Note that $N^2=0$ at the disk midplane.  In
general
it will become large far from the midplane.  This dispersion relation implies
that $N^2(z)<\bar\omega^2<\Omega^2$.  Since $\bar\omega$ is a function of
$r$ for non-axisymmetric waves this implies that internal waves are confined
vertically and radially.  The vertical trapping concentrates these waves in
regions of low $N^2$ (like the midplane).  One of the radial boundaries
is defined by $\bar\omega^2=\Omega^2$, where the wave undergoes reflection.
At the other boundary, as $\bar\omega\rightarrow 0$, $k_r\rightarrow\infty$.
The wave slows down without reflecting and its energy becomes concentrated
towards
the disk midplane and the radial boundary.  In practice this implies nonlinear
dissipation of the wave energy, a phenomenon referred to as resonant
absorption.
The existence of an absorbing boundary would seem to make such waves distinctly
unpromising as a means of long range transport, but there is one important
exception.
A slightly non-axisymmetric internal wave, $|m|=1$, with a small $\omega$ can
have a resonant absorption radius far outside the disk and a reflecting
boundary
at $r=0$. For example, in a binary system such waves could be excited
either by the impact of the accretion stream or by tidal forces with a
frequency
of $\Omega_b$, which is necessarily less than $\Omega$ at the outer edge of the
disk.  Recently, Goodman (1993) has pointed out that there is a generic
parametric
instability driven by tidal forces which will produce exactly this kind of
wave.
Of course, $m=0$ waves can also move within the disk without encountering
reflecting or absorbing radial boundaries.  However, since these waves are
axisymmetric they do not carry angular momentum and cannot twist up
magnetic fields (so as to cause dynamo action).  Furthermore, a low frequency
wave propagating inward will necessarily become increasingly confined to
the disk midplane.

It's worth noting that in addition to their other virtues, $m=1$ internal
waves undergo linear amplification as they propagate inward.  This comes
about because their total wave energy flux is the sum of two parts, i.e.
\begin{equation}
F_r=\langle \delta P v_r\rangle +\Omega {\cal L}_r,
\end{equation}
where ${\cal L}$ is the angular momentum flux.  Both $F_r$ and ${\cal L}_r$
are conserved to linear order.  An inward moving wave with $m=1$
will have ${\cal L}_r$ positive but $\langle \delta P v_r\rangle$
negative so that both increase together as $r\rightarrow 0$ and
$\Omega\rightarrow \infty$.  Consequently, $\langle v^2\rangle$
also increases and the waves grow to a limit imposed by nonlinear
processes.  The growth rate is just $V_{group}/r\sim (H/r) \Omega$,
the rate at which the waves travel a significant radial distance.

Convection is closely related to internal waves, in the sense that they
appear in Eq. (\ref{eq:disp}) for $N^2<0$.  For $\Gamma_{conv}\equiv|N|$
small compared to $\Omega$ typical convective modes have
$k_r\sim (\Omega/\Gamma_{conv})k_z$ and $m\sim rk_z$.  The angular
momentum transport induced by such cells can be as large as
an `$\alpha$' of $(\Gamma_{conv}/\Omega)^3$ (assuming $k_z\sim H^{-1}$).
This will not be important unless $\alpha$ is otherwise quite small or
convection is quite strong.  Worse, the sign of this effect is not yet
known.  Ryu \& Goodman (1992) have argued that the dominant modes
transport angular momentum {\it inward}.  Lin, Papaloizou \& Kley (1993)
have argued
that modes with transport effects of either sign must occur and that
correctly accounting for boundary conditions will inevitably lead to
the outward transport of angular momentum.  Diamond, Vishniac \& Luo
(1993) have demonstrated that although modes with both signs of
angular momentum transport do occur, they differ in their systematic
properties in ways that suggest that the inward transport of angular
momentum may be the net effect.  The conservative position is probably to
demand a fully nonlinear treatment of the problem.

In addition to these hydrodynamic modes, the presence of a large scale
azimuthal
magnetic field implies the possibility of Alfv\'en waves and related modes.
The most important of these are long wavelength (i.e. $\lambda>V_A/\Omega$)
instabilities related to radially polarized Alfv\'en waves.  This
instability was described in the context of couette flow more than 30 years
ago (Velikhov 1959, Chandrasekhar 1961), but its significance for accretion
disks has only been recognized recently (Balbus \& Hawley 1991).
It has usually been described in terms of a vertical magnetic field, but also
occurs
for an azimuthal field.  It appears whenever $\partial_r\Omega\ne 0$, which
is necessarily true in a thin accretion disk.  Moreover, it automatically
moves angular momentum so as to minimize the gradient in $\Omega$ (rather
than in $r^2\Omega$).  Consequently it is guaranteed to move angular
momentum outward in an accretion disk, a necessary part of any successful
model.  The growth rate is roughly $mV_A/r$ for small $m$, but cuts off at
a maximum value somewhat less than $\Omega$ on a scale of $V_A/\Omega$.
On smaller scales it disappears altogether.  The instability saturates
in a turbulent state with velocities of order $V_A$ on scales of $V_A/\Omega$.
Smaller $m$ modes are largely suppressed by the small scale turbulence so
that the turbulent diffusion can be approximately modeled by a diffusion
coefficient of $D\sim V_A^2/\Omega$.  Similarly, the angular momentum can
be modeled by an effective $\alpha\sim (V_A/c_s)^2$.
Although the driving mechanism for this instability comes
from radial motions, shearing stresses and the condition of approximate
incompressibility lead to approximately isotropic turbulence.
There is now widespread agreement that this modes are critically important
for angular momentum transport in accretion disks.  However, an
important question is how the magnetic field necessary to
drive the Velikhov-Chandrasekhar instability is maintained.

One traditional objection to the presence of significant magnetic fields
in accretion disks is that buoyancy will tend to eject any magnetic
field on a time scale much shorter than radial infall time.  This comes
about because magnetic fields supply pressure, but not mass, so that
flux tubes tend to be buoyant.  In practice the dominant buoyant mode
involves vertical ripples which allow the matter tied to the field lines
to sink while the magnetic field forms rising bubbles.  These bubbles
will almost certainly reconnect as they rise, leading to a net loss of
flux.  This is the Parker instability (Parker 1971, 1979) and while it can be
suppressed in some circumstances
it is quite difficult to avoid in general. Even if the entropy gradient
of the gas leads to a stabilization, small conductive effects will still
allow this process to continue, albeit at a reduced rate.  These modes
can be thought of as long azimuthal wavelength, vertically polarized
Alfv\'en waves.  The critical modes have azimuthal wavelengths comparable
to the height of the disk and grow at a rate of $V_A/H$.  If this process
were effective it would lead to a flux loss rate of approximately
$V_A/H$.  However, this process is disrupted by the Velikhov-Chandrasekhar
instability, which necessarily accompanies the existence of an azimuthal
field.  This follows from a simple qualitative argument.  An instability
in an accretion disk with a growth rate $\Gamma$ has a radial wavenumber
constrained by the condition that
\begin{equation}
k_r>{\partial_r\bar\omega\over\Gamma}\sim {m\Omega\over r\Gamma},
\end{equation}
since smaller wavenumbers will get sheared out in less than one e-folding
time of the instability.  The Parker instability has $m\sim r/H$ so
\begin{equation}
k_r>{\Omega\over V_A}.
\end{equation}
On such radial scales the Velikhov-Chandrasekhar instability will mix the
vertical momentum of rising and falling flux tubes at a rate $\sim \Omega$,
i.e. much faster than the growth time of the Parker instability.  Of course,
the buoyancy of the field lines cannot be entirely suppressed, but the speed
with which flux tubes can rise will be reduced by a factor of $\Gamma/\Omega$
so that the magnetic flux loss rate from a disk will be of order
$(V_A/c_s)^2\Omega$.  This is approximately what one would expect from
turbulent
diffusion due to the Velikhov-Chandrasekhar instability, the only difference
being that the buoyant motions have a preferred direction.  In other words,
the Parker instability will operate at greatly reduced efficiency if $V_A<c_s$,
and the magnetic flux loss rate will be roughly what is expected from turbulent
diffusion.

I note in passing that the buoyant rise of an axisymmetric flux tube (cf.
Sakimoto \& Coroniti 1989) will still involve the turbulent entrainment of
the neighboring fluid due to the Velikhov-Chandrasekhar instability so that
the magnetic flux loss rate is given by the preceding estimate
(Vishniac 1993).

Applying similar considerations to the theory of convection in a magnetized
disk, we find that a sufficiently powerful magnetic field will disrupt the
convective cells.  Assuming that the circulation within the cells will create a
locally ordered field whose strength is just sufficient to modify the nature of
the convective flow, one finds that the angular momentum transport associated
with the Velikhov-Chandrasekhar instability of this field is comparable in
magnitude to the angular momentum transport associated with purely
hydrodynamic convection.  The only qualitative difference is that this
contribution is guaranteed to be positive.  Evidently a complete theory
of convection in cataclysmic variable disks must include some consideration
of the magnetic field swept up in the convective motions.

\section{The Internal-Wave Driven Dynamo}

The idea behind the internal-wave driven dynamo model for angular momentum
transport is to combine these perturbative modes with a dynamo model in
order to produce a complete model of accretion disk dynamics.  The
chain of causation in this model runs from the tidal forces that excite
$m=1$ internal waves, through the waves themselves and the processes
that determine their amplitude as a function of radius, including the
generation of subharmonics by nonlinear wave interactions, through a
magnetic dynamo mediated by the full spectrum of internal waves, and culminates
in the appearance of magnetic field instabilities which saturate the
dynamo action and induce positive angular momentum transport.
A flow chart for this model looks like this:
\vskip 1cm
\centerline{EXCITATION OF INTERNAL WAVES - tidal forces cf. Goodman}
\vskip 0.3cm
\hskip 5.75cm $\Big\downarrow$\ (+ Linear Amplification)
\vskip 0.3cm
\hskip 5.75cm $\Big\downarrow$\ (+ Nonlinear Wave Interactions)
\vskip 0.3cm
\centerline{SATURATED WAVE SPECTRUM}
\vskip 0.3cm
\hskip 5.75cm $\Big\downarrow$\ (+ Shearing)
\vskip 0.3cm
\centerline{$\alpha-\Omega$ DYNAMO}
\vskip 0.3cm
\hskip 5.75cm $\Big\downarrow$
\vskip 0.3cm
\centerline{GROWTH OF $B_\theta$}
\vskip 0.3cm
\hskip 5.75cm $\Big\downarrow$\ (+ VC Instability)
\vskip 0.3cm
\centerline{SMALL SCALE TURBULENCE}
\vskip 0.3cm
\hskip 5.75cm $\Big\downarrow$
\vskip 0.05cm
\noindent\hskip 0.9cm\vbox{\hrule width9.7cm}
\vskip 0.05cm
$\Big\downarrow$\hskip 3.0cm $\Big\downarrow$\hskip 3cm $\Big\downarrow$\hskip
3cm $\Big\downarrow$
\vskip 0.3cm
\noindent SATURATION\hskip 0.4cm TURBULENT\hskip 0.6cm TRUNCATION\hskip 0.6cm
SUPPRESSION
\vskip 0.1cm
OF $B_{\theta}$\hskip 1.1cm TRANSPORT\hskip 0.7cm OF INTERNAL\hskip 0.5cm OF
PARKER
\vskip 0.1cm
\noindent\hskip 6.9cm WAVE\hskip 1.4cm INSTABILITY
\vskip 0.1cm
\noindent\hskip 6.4cm SPECTRUM
\vskip 1cm
I've already alluded to most of the steps in this process.  The major gap lies
in the discussion of the dynamo itself.  Typical proposals for disk dynamos
are turbulent $\alpha-\Omega$ dynamos.  In this kind of scheme an azimuthal
field is acted on by some local, turbulent velocity field so that the
unperturbed
field line is replaced by a spirally flux tube centered on the original field
line position.  If a vertical stack of such flux tubes then undergo
reconnection
a vertical gradient in field strength (or the amplitude of the velocity
field) results in the production of radial flux.  The differential shearing of
the
disk results in a growth rate for $B_\theta$ of $-(3/2)\Omega B_r$, thereby
closing the loop.  The growth rate for $B_r$ is
$-\partial_z \alpha_{\theta\theta}B_\theta$ (neglecting dissipative terms),
where
\begin{equation}
\alpha_{\theta\theta}=\langle v_z{1\over r}\partial \int^t v_r(t')dt'
-v_r{1\over r}\partial \int^t v_z(t')dt'.\rangle
\end{equation}
Note that an isotropic velocity field will give $\alpha_{\theta\theta}=0$,
even if the disk turbulence is quite strong.  Moreover, whatever
asymmetry is present must be quite strong since the velocity field
will also tend to have a dissipative effect.

In the wave-driven dynamo model the turbulence is replaced by an
ensemble of internal waves, consisting of the $m=1$ waves that
are generated in the outer parts of the disk, and the higher order
waves generated locally through nonlinear interactions.  Waves of
infinitesimal amplitude will have $\int^t \vec v dt'=0$, since they
can be represented as the sum of perfectly periodic motions.
However, if the waves are balanced between nonlinear dissipation
and linear amplification then they will not be periodic, but
suffer from a slight loss of coherence due to nonlinear effects.
Such an effect is roughly analogous to the collisional broadening of
spectral lines.  In this case the decoherence rate is equal to the
linear amplification rate, $(H/r)\Omega$.  Consequently the
time integral becomes
$\int^t\vec v dt'\approx \vec v (\bar\omega^2\tau_{decoherence})^{-1}$.
The necessary asymmetry in the velocity field is supplied by the
dominance of ingoing waves.  This dominance is guaranteed by the
fact that the ingoing waves amplify and the excitation region for the
waves is near the outer edge of the disk.  In fact, even
in the outer parts of the disk there will be such an asymmetry since
the growth rate for the tidal instability is strong function of radius
(Goodman 1993).  One other distinctive feature of this kind of dynamo
is that the waves that drive it have long wavelength symmetries.  Unlike
typical turbulent dynamos the wave-driven dynamo involves motions with
vertical and radial wavelengths comparable to the disk thickness, and
an azimuthal wavelength of $2\pi r$.  The emergence of large scale
structure in the magnetic field is less mysterious under such circumstances.

The growth rate of the dynamo is roughly
$(\alpha_{\theta\theta}\Omega/H)^{1/2}$.
If we are far enough from the disk edge that the tidal instability can
be neglected then $\tau_{decoherence}^{-1}\sim (H/r)\Omega$.  Since
the internal wave dispersion relation gives a group velocity which is
a strong function of $\bar\omega$ it is reasonable to invoke an
incoherent nonlinear damping rate, i.e.
\begin{equation}
\tau_{decoherence}^{-1}\sim {\langle v_z^2\rangle\over\bar\omega H^2}.
\end{equation}
Consequently, $\langle v^2\rangle\sim (H/r)c_s^2$.  Folding these together
with the definition of $\alpha_{\theta\theta}$ we find that in the
inner parts of the disk the dynamo growth rate {\it due to the $m=1$
waves} will be approximately $(H/r)^{3/2}\Omega$.  However, note that
nonlinear interactions between these waves will generate modes with
lower $\bar\omega$ and higher $m$.  Consequently, the contribution
to $\alpha_{\theta\theta}$ from subharmonics may very well dominate
provided that the asymmetry in the wave distribution is preserved during
the nonlinear cascade.   Preliminary work by Huang (1992) indicates
that it is preserved and an estimate of its effects (Vishniac and
Diamond, 1992) suggests that the total dynamo growth rate may
be as high as $\sim (H/r)^{4/3}\Omega$.  This implies an equivalent
`$\alpha$' of $(H/r)^{4/3}$.  Towards the outer parts of the disk
the angular momentum transport may be higher, due to the direct
excitation of internal waves from tidal instabilities.

Ultimately the contribution from subharmonics is limited by the
disruptive effect of the small scale turbulence due to the
Velikhov-Chandrasekhar instability.  As one goes down the
weakly turbulent cascade towards small $\bar\omega$ the wave-wave
interactions are eventually overwhelmed by the dissipative effects
of this shearing instability.  This effect provides the stabilizing
feedback for the whole system.  A stronger magnetic field will
act to disrupt a larger part of the internal wave distribution, lowering
$\alpha_{\theta\theta}$ and the dynamo growth rate.  Meanwhile, this
same turbulence will increase the magnetic flux loss rate from the
disk.  A weakened magnetic field will have the opposite effect, resulting
in a growth of the mean magnetic field strength.  At equilibrium the
local magnetic field will have a disordered component roughly as strong
as the long range component.  The average Alfv\'en speed will be
$\sim \alpha^{1/2}c_s\sim (H/r)^{2/3} c_s$.  The typical turbulent
eddy size will be $\alpha^{1/2} H\sim (H/r)^{2/3}H$.  It follows that
in thin disks the dissipation of orbital energy will occur on scales
much smaller than the disk thickness.

\section{Observations and Possible Tests}

In its present form the internal wave-driven dynamo model can
only yield a set of scaling laws.  In spite of this there are
several points of contact with the observations, and as
the model becomes more quantitative we can expect decisive
observational tests of the model.  Here I will note only
a few of the more interesting possible tests.

First, in this model thicker disks imply larger dimensionless
viscosities.  This should be directly applicable to stationary
disks (where the truth of this assertion is hard to judge) and
could be incorporated into models of dwarf novae outbursts.  Such
models typically suggest a larger `$\alpha$' during outbursts
(e.g. Mineshige \& Osaki 1983, 1985, Smak 1984, Meyer 1984, Cannizzo,
Wheeler \& Polidan 1986),
but the nonlocal nature of $\alpha$ in this model
implies that a detailed implementation of this model is necessary
before any conclusions can be drawn.

Second, nonthermal heating in a disk atmosphere can come from
the dissipation of weak shocks generated within the disk, the
eruption of magnetized bubbles of plasma, and the appearance of
turbulent eddys whose size (relative to $H$) is not small.
All of these effects occur in this model.  The first involves
the diversion of some fraction of the energy in the weakly
turbulent wave cascade into compressive modes.  The total
flux from this will scale as $\langle v^2\rangle\tau^{-1}$
or $(H/r)^2$.  This implies that the fraction of the local heating
budget released in this way is $\sim (H/r)^{2/3}$.  The
flux of magnetic energy will be proportional to $V_A^2 (V_A/c_s)^2c_s$,
which is a fraction $\sim (H/r)^{4/3}$ of the total energy.
The last contribution is dependent on the local structure of the
disk, but will certainly increase with $V_A/c_s$.  It follows
that coronal heating will be a function of the disk
geometry, with fatter disks showing stronger coronae.  If the
fractional {\it local} coronal heating rate can be measured, then it should
increase with $r$ (since $H/r$ typically does).

Third, in some systems (like MV Lyrae or TT Ari) the mass flow is
intermittent.  When such a system is in an extreme low state the
outer edge of the disk should gradually shrink, reducing the
role of the tidal excitation of internal waves and thereby
diminishing the value of $\alpha$ in the remaining disk.  The
dynamics of such disks should therefore reflect the persistence of
a relatively inactive disk, even after long periods of very low
mass flux.

Fourth, instabilities in radiation dominated disks should be
profoundly affected by the non-local nature of angular momentum
transport (cf. Vishniac 1993).  This may be difficult to check
using AGN, where the observations are somewhat ambiguous, but
disks around galactic black holes may provide useful observational
clues.

\section{Discussion}

This is an active field, and the details presented here are
likely to change in the near future.  However, it seems likely
that the basic points listed below will remain a part of any
viable model of angular momentum transport in accretion disks.

First, the Velikhov-Chandraskhar instability is the {\it only}
kind of turbulent stirring guaranteed to produce outward transport
of angular momentum in non-selfgravitating disks.  Turbulence
due to other instabilities (e.g. convection) may transport angular
momentum in either direction.  The basic difficulty here is that
pure mixing processes will tend to equalize conserved quantities
per particle, like angular momentum.  In an accretion disk this will
drive angular momentum inward.  This means that angular momentum
transport in magnetized disks is almost certain to involve the
Velikhov-Chandrasekhar instability.  Transport in neutral, highly
resistive, disks is apt to be much less efficient.

Second, convection is important only when it is strong, but the
sign of $\alpha_{conv}$ is unknown.  In a magnetized disk it is
apt to be more positive than in a neutral disk and its effects
will not be confined to the convectively unstable layer.

Third, magnetic flux escapes from a disk at a rate
$\sim (V_A/c_s)^2\Omega\sim \alpha_{VC}\Omega$.  This is substantially
slower than previous estimates due to the turbulent mixing of the
gas caused by the Velikhov-Chandrasekhar instability.  Balancing this
loss rate with a dynamo implies $\Gamma_{dynamo}\sim \alpha_{VC}\Omega$.

Fourth, these points imply that a local dynamo gives an $\alpha$ which
is a universal constant.  There is, as yet, no theoretical reason
to believe that such a dynamo emerges from the Velikhov-Chandrasekhar
instability (or any other local process), but it is difficult to
rule out this possibility.  Phenomenological arguments tend to
lead to a rejection of this possibility, but the role of convection
in real objects could invalidate the models used to date.

Fifth, the internal wave-driven dynamo gives an `$\alpha$' which is
non-local and scales as $(H/r)^{4/3}$ in a stationary disk.  The
dynamics of such a model are not yet known.

Sixth, excitation of the internal waves in a binary system will
be mostly due to a tidal instability.  The efficiency of this
process will be a fairly strong function of radius.  The presence
of tidal resonances will reinforce this process.  Systems with
moderate mass ratios and intermittent mass flows may shrink and
become cold during extended interruptions of the mass flow.

Seventh, the Velikhov-Chandrasekhar instability will produce
strong vertical mixing, with an effective vertical diffusion
coefficient of $\alpha Hc_s$.  To see this one can estimate the
heat flux due to vertical mixing as
\begin{equation}
F_{mixing}\sim P {D_z\over L_S},
\end{equation}
where $L_S$ is the vertical entropy length scale and I have
assumed that the fluid is optically thick.  Since
$D_z\sim \alpha Hc_s$ and the radiative flux is
\begin{equation}
F_{radiative}\sim \dot M\Omega^2\sim \alpha Pc_s.
\end{equation}
It follows that
\begin{equation}
F_{mixing}\sim F_{radiative} {H\over L_S}.
\end{equation}
In other words, in an optically thick disk with a significant
vertical entropy gradient the turbulent mixing
will lead to a heat flow which is comparable to the radiative
flux, but can have either sign.  This will tend to stabilize
disk models, since the vertical structure will no longer be as
sensitive to the details of the opacity, and reduce the magnitude
of the vertical entropy gradients in such models.  A simple
approximation of this effect could, and should, be incorporated
into current models of vertical structure.

\acknowledgments

The work presented here is the result of an ongoing collaboration
involving several other researchers, principally P.H. Diamond.
L. Jin, M. Huang, S. Luo,
and W. Zhang are also responsible for some of the results summarized
here.  This work was supported by NASA through research grant NAGW-2418.

\end{document}